\documentclass{svjour2}                    

\smartqed  
\usepackage{graphicx}
\usepackage{color}

\usepackage{amssymb}
\begin{document}

\title{Parallel tempering Monte Carlo simulations of spherical fixed-connectivity model for polymerized membranes
}

\titlerunning{Parallel tempering Monte Carlo simulations of spherical fixed-connectivity model}        

\author{Satoshi Usui and Hiroshi Koibuchi}


\institute{
Satoshi Usui \at
Advanced Course of Mechanical Engineering, National Institute of Technology, Ibaraki College, Nakane 866, Hitachinaka, Ibaraki 312-8508, Japan
 \\
Hiroshi Koibuchi \at
Department of Mechanical and Systems Engineering, National Institute of Technology, Ibaraki College, Nakane 866, Hitachinaka, Ibaraki 312-8508, Japan
 \\
              \email{koibuchih@gmail.com}          
}


\maketitle

\begin{abstract}
We study the first order phase transition of the fixed-connectivity triangulated surface model using the Parallel Tempering Monte Carlo (PTMC) technique on relatively large lattices. From the PTMC results, we find that the transition is considerably stronger than the reported ones predicted by the conventional Metropolis MC (MMC) technique and the flat histogram MC technique. We also confirm that the results of the PTMC on relatively smaller lattices are in good agreement with those known results. This implies that the PTMC is successfully used to simulate the first order phase transitions.  The parallel computation in the PTMC is implemented by OpenMP, where the speed of the PTMC on multi-core CPUs is considerably faster than that on the single-core CPUs. 

\keywords{Triangulated surfaces \and Phase transition \and Surface fluctuation \and Parallel tempering Monte Carlo}
 \PACS{ 64.60.-i  \and 68.60.-p  \and 87.16.D-}
\end{abstract}
\section{Introduction}
The shape transformation of giant vesicles is very interesting. This transformation can also be seen in biological membranes such as living cells \cite{NELSON-SMMS2004-Leibler}. Since the membranes are very soft, their shape changes even with small forces such as the thermal fluctuation \cite{NELSON-SMMS2004-Gompper,Bowick-PREP2001,Wiese-PTCP2000}. To understand the membrane fluctuations and shape transformations a lot of studies have been conducted \cite{KANTOR-NELSON-PRA1987,Peliti-Leibler-PRL1985,DavidGuitter-EPL1988,PKN-PRL1988,BKS-PLA2000,Kownacki-Mouhanna-2009PRE}. For a statistical mechanical model of membranes, a finite upper critical dimension $d_c$, above which the Landau theory holds, is expected from the mean field approximation \cite{PKN-PRL1988}. Recently, it is predicted that $d_c\simeq 5$ from a renormalization group study \cite{Essa-Kow-Mouh-PRE2014}. 

Numerically, it was reported that the triangulated surface model undergoes a first order transition in $d\!=\!3$ \cite{KD-PRE2002}. The first order transition can also be seen in the canonical surface model \cite{KOIB-PRE-2005,KOIB-JSTP-2010}, which was introduced by Helfrich and Polyakov \cite{HELFRICH-1973,POLYAKOV-NPB1986}, and the same transition is expected in the Landau-Ginzburg model for membranes \cite{KOIB-SHOB-IJMPC-2014}.  All results in those studies support that the surface model in $d\!=\!3$ undergoes a first order phase transition between the smooth and crumpled phases at relatively small bending rigidity.  However, the simulations of the surface models are known as very time consuming at least for the Monte Carlo (MC) techniques. In fact, the conventional MC technique, such as the Metropolis MC (MMC) technique \cite{Mepropolis-JCP-1953,Landau-PRB1976}, is designed for the single CPU computers, and therefore the MMC technique is not always efficient for simulations on the current multi-core CPUs. Therefore it is necessary and worthwhile to study how to accelerate the speed of the surface simulations using the currently available multi-core CPUs. 

The Parallel Tempering Monte Carlo (PTMC) technique is originally constructed for the spin glass simulations \cite{Hukushima-Nemoto-JPSOC1996,Takayama-Hukushima-JPSOC2007}. At the low temperatures, the acceptance rate of the spin flips is very low in the conventional MC techniques including the heat bath MC. This slow convergence is greatly improved by PTMC, and therefore this technique is successfully applied to study the spin glass transitions.  However, the problem is whether or not PTMC can be used to simulate the first order phase transitions \cite{Hukushima-Nemoto-JPSOC1996}. On this problem a lot of studies have been conducted, and currently it is believed that PTMC can be applied to the simulations of the first order transitions \cite{New-Mag-Hans-PRE2007,C-E-Fios-PRE2008}. 

However, it is still unclear whether or not the PTMC technique is applicable to the first order transition of the fixed-connectivity surface model. In this paper, we study the fixed-connectivity surface model by means of PTMC with the OpenMP parallelization,  which is well known as an easy-to-use technique and efficient for the multi-core CPUs \cite{Intel}. The main purpose of this study is to find that the PTMC is a correct simulation technique for the first order transition of the surface model. Indeed, it is possible to check whether or not the first order transition of the model remains unchanged on the lattices which are considerably larger than those used up to now. Moreover, it is practically interesting to see whether or not the PTMC simulation is efficient (or the speed is very fast) for the membrane simulations. 

\section{Model}\label{model}
The Hamiltonian of the model is defined on the triangulated lattices of sphere topology. The vertex position is denoted by ${\bf r}_i (\in {\bf R}^3)$, $(i=1,\cdots,N)$, where $N$ is the total number of vertices. The total number of bonds (edges of triangles) is given by $N_B\!=\!3N\!-\!6$, and the total number of triangles is $N_T\!=\!2N\!-\!4$.  

The partition function $Z$ of the model is defined by \cite{AMBJORN-NPB1993,WHEATER-JP1994}
\begin{eqnarray} 
\label{part-func}
Z(\kappa) = \int^\prime \prod _{i=1}^{N} d {\bf r}_i \exp\left[-S({\bf r}; \kappa)\right],
\end{eqnarray} 
where the prime in the integral $\int^\prime \prod _{i=1}^{N} d {\bf r}_i$ denotes that the center of mass of the surface is fixed to the origin of ${\bf R}^3$ to protect the surface from the translation. The Hamiltonian $S({\bf r}; \kappa)$ is given by 
\begin{eqnarray}
\label{Hamiltonian} 
&& S({\bf r}; \kappa)=S_1 + \kappa S_2,  \nonumber \\
&& S_1=\sum_{ij} \left( {\bf r}_i-{\bf r}_j\right)^2, \quad S_2=\sum_{ij} (1-{\bf n}_i \cdot {\bf n}_j), 
\end{eqnarray} 
where $S_1$ and $S_2$ are the Gaussian bond potential and the bending energy, respectively.  The symbols $\kappa[1/k_BT]$ and ${\bf n}_i$ are respectively the bending rigidity and the unit normal vector of the triangle $i$, where $k_B$ and $T$ are the Boltzmann constant and the temperature. 

\section{Parallel tempering  Monte Carlo}

In MMC, the distribution of the vertex position ${\bf r}$ becomes proportional to $\exp(-S)$ \cite{Mepropolis-JCP-1953}. Indeed, ${\bf r}$ is moved to a new position ${\bf r}^\prime\!=\!{\bf r}\!+\!\delta{\bf r}$ with the  probability ${\rm Min}[1,\exp(-\Delta S)]$, where $\delta S$ is the energy change $\Delta S\!=\!S({\bf r}^\prime)-S({\bf r})$ before and after the vertex move. As a consequence, the probability for the positional changes from ${\bf r}$ to ${\bf r}^\prime$ and ${\bf r}^\prime$  to ${\bf r}$  satisfies the detailed balance condition,  and the distribution of $\{{\bf r}\}=\{{\bf r}_1, {\bf r}_2, \cdots, {\bf r}_N\}$ becomes proportional to $\exp(-S)$ \cite{Landau-PRB1976,Karin-Taylor-1975,A-Ueda-1990}. However, the MMC technique is not always effective for the multi-core CPUs because it is originally designed for the single-core CPUs.

Let $N_R$ denote the total number of replicas  \cite{Hukushima-Nemoto-JPSOC1996}. In PTMC, these $N_R$ replicas are simulated in a single simulation.  Each system is in the one-to-one correspondence to the temperature, which belongs to the region $[T_{\rm min},T_{\rm max}]$, and this correspondence is exchanged during the simulation. This exchange process is the essential part of PTMC. Note that the total number of temperatures to be simulated is identical with the total number of replicas $N_R$.

In the surface model for membranes, the parameter is the bending rigidity $\kappa$, which plays a role of the inverse temperature $1/T$. 
Let $\{{\bf r},\kappa \}$ stand for $N_R$ replicas $\{{\bf r}_1,\kappa_1; {\bf r}_2,\kappa_2; \cdots; {\bf r}_{N_R},\kappa_{N_R}\}$. We assume that the increment $d\kappa$ of two neighboring $\kappa$ is uniformly given by $d\kappa\!=\!(\kappa_{\rm max}\!-\!\kappa_{\rm min})/N_R$, where $\kappa_{\rm max}$ and $\kappa_{\rm min}$ are the maximum and minimum bending rigidities corresponding to $T_{\rm max}$ and $T_{\rm min}$ in the above mentioned case for spin glasses.  
 
The basic processes of PTMC to be performed for the surface model are as follows \cite{Hukushima-Nemoto-JPSOC1996}:
\begin{enumerate}
\item[(i)] Perform  $N_s$ MCS (Monte Carlo sweeps) for each system $({\bf r},\kappa)$
\item[(ii)] Exchange the systems  $({\bf r},\kappa)$  and $({\bf r}^\prime,\kappa^\prime)$ in the nearest neighboring  $\kappa, \kappa^\prime$ with the probability
\begin{eqnarray}
\label{dbal-exchg-1}
&&W({\bf r},\kappa|{\bf r}^\prime,\kappa^\prime)={\rm Min}\left[1,\exp(-\Delta)\right], \nonumber \\
&&\Delta=(\kappa^\prime-\kappa)\left[S_2({\bf r})-S_2({\bf r}^\prime)\right] 
\end{eqnarray}

\item[(iii)] Repeat the processes (i) and (ii)
\end{enumerate}
We remark that Eq. (\ref{dbal-exchg-1}) implies that the systems $({\bf r},\kappa)$  and $({\bf r}^\prime,\kappa^\prime)$ should be exchanged with the probability 1 if $\Delta$ is smaller than 0, and with the probability $\exp(-\Delta)$ if $\Delta$ is larger than 0. It should also be remarked that $\Delta$ in Eq. (\ref{dbal-exchg-1}) is given only by the difference of $S_2$ and includes no information on the difference of $S_1$. In the process (ii), every system $({\bf r},\kappa)$ is updated once. The acceptance rate depends on $d\kappa$, which should be fixed sufficiently small.

The systems are all independently updated in the process (i). The total number of MCS in (i) is $N_s\!=\!20$ for small lattices and $N_s\!=\!30$ for larger lattices. Thus, the process (i) becomes the most time consuming part of the simulation, and therefore the process (i) should be parallelized and this parallelization is expected to be effective. In the process (i), the conventional MMC technique is employed. In this MMC technique, it is well known that the vertex position is updated so that the probability of the change satisfies the detailed balance condition \cite{Karin-Taylor-1975,A-Ueda-1990}. 
We call one iteration of (i) and (ii) as 1 MCS of the PTMC; $N_s$ MCS of MMC in (i) is included in 1 MCS of the PTMC. 

The problem is whether the processes (i) and (ii) produce the canonical distribution of the configuration for every replica $({\bf r},\kappa)$. In fact, the surface model is not always exactly corresponding to the ferromagnetic or spin glass model, because the phase space of the surface model is ${\bf R}^3$, which is quite different from those for spin models. Therefore we should comment on this problem. First of all, it is clear that the process (i) simply produces the canonical configuration with the weight $\exp[-S({\bf r}_i,\kappa_i)]$ for each $({\bf r}_i,\kappa_i)$, because every system is independently  updated with a given $\kappa_i$. However, the bending rigidity $\kappa$ (or ${\bf r}$) is exchanged in the process (ii), and therefore it must be clarified that the process (ii) does not violate the canonical distribution as a result of the process (i). 

Let ${\mathcal C}$ be the set of replicas  $\{{\bf r},\kappa\}$ for all possible exchanges of $\kappa$. The total number of the systems in ${\mathcal C}$ is $N_R !$. Let $\sigma_i$ denote the element of ${\mathcal C}$ such that ${\mathcal C}\!=\!\{\sigma_1, \sigma_2,\cdots, \sigma_{N_R!}\}$, and $\sigma_i$ corresponds to a replica $({\bf r}_i,\kappa_i)$. This set ${\mathcal C}$ is the configuration space in which the micro-state is wandering in the exchange process. Indeed, the set of the configuration $\{{\bf r}_1, {\bf r}_2,\cdots, {\bf r}_{N_R}\}$ remain unchanged and only the combination of ${\bf r}$ and $\kappa$ is replaced in the exchange process. Thus the combination of ${\bf r}$ and $\kappa$ is summed over in the partition function and dynamically changed for the extended system.

The basic assumption for the PTMC is that the exchange process is Markovian and ergodic. Under these conditions, it is well known that if the detailed balance condition for the exchange probability $P$ is satisfied then there uniquely exists the limiting distribution of $P$ for any state $\sigma$ in ${\mathcal C}$. 
 Here we should note that 
 the probability distribution $P$ of the variable $\sigma (\in {\mathcal C})\!=\!\{{\bf r},\kappa \}\!=\!\{{\bf r}_1,\kappa_1; {\bf r}_2,\kappa_2; \cdots, {\bf r}_{N_R},\kappa_{N_R}\}$ is given by
\begin{eqnarray}
\label{dbal-exchg-3}
&&P(\{{\bf r},\kappa \})=\prod_{m=1}^{N_R}P_{\rm eq}({\bf r}_m,\kappa_m), \nonumber \\
&&P_{\rm eq}({\bf r},\kappa)=Z^{-1}(\kappa)\exp\left[-S({\bf r},\kappa)\right], \nonumber \\
&&S({\bf r},\kappa)=S_1({\bf r})+\kappa S_2({\bf r}).
\end{eqnarray}
This can be assumed independently of the process (i), because Eq. (\ref{dbal-exchg-1}) is used for the exchange process.

Thus, the only task we should see is that the PTMC exchange satisfies the detailed balance condition, which is  described as
\begin{eqnarray}
\label{dbal-exchg-2}
&&P(\cdots; {\bf r},\kappa; \cdots; {\bf r}^\prime,\kappa^\prime;\cdots)W({\bf r},\kappa|{\bf r}^\prime,\kappa^\prime) \nonumber \\
&&=P(\cdots; {\bf r}^\prime,\kappa; \cdots; {\bf r}^\prime,\kappa;\cdots)W({\bf r}^\prime,\kappa|{\bf r},\kappa^\prime), 
\end{eqnarray}
where $P$ is the probability distribution function used in Eq. (\ref{dbal-exchg-3}), and $W$ is the exchange probability defined in Eq. (\ref{dbal-exchg-1}). The symbol ${\bf r},\kappa$ in $P$ and $W$ is an abbreviation of $({\bf r}_1,\cdots, {\bf r}_N;\kappa)$, which denotes a replica with the bending rigidity $\kappa$ just the same as above. To show that the canonical distribution of $P$ is consistent with the exchange process (ii), it is sufficient to prove that the detailed balance condition of Eq. (\ref{dbal-exchg-2}) is equivalent with 
\begin{eqnarray}
\label{dbal-exchg-4}
W({\bf r},\kappa|{\bf r}^\prime,\kappa^\prime) / W({\bf r}^\prime,\kappa|{\bf r},\kappa^\prime)=\exp(-\Delta)
\end{eqnarray}
 under the condition of Eq. (\ref{dbal-exchg-3}). Indeed, this equivalence is straight forward to prove.
From this and the above mentioned uniqueness theorem, it is easy to understand that the function $P(\sigma)$ in Eq. (\ref{dbal-exchg-3}) is the unique distribution for arbitrary micro-state  $\sigma (\in {\mathcal C})$ if the exchange in the process (ii) is updated according to  Eq. (\ref{dbal-exchg-1}). 

We should note that Boltzmann distribution  for ${\bf r}_i$ can also be assumed, where ${\bf r}_i$ represents $N$ sets of vertex position $\{{\bf r}_1, {\bf r}_2,\cdots, {\bf r}_N\}$ corresponding to $\kappa_i$. Indeed, this assumption is expected to be always satisfied, because the canonical distribution of the configuration ${\bf r}_i$ corresponding to $\kappa_i$ is generated in the process (i) of the PTMC simulation. In the process (ii), as described above, the state $\sigma (\in {\mathcal C})$ is expected to satisfy the canonical distribution. This makes us to calculate the physical quantities using arbitrary element $\sigma$. 

The next and main problem to be clarified in this paper is whether or not the PTMC technique can be used to simulate the first order phase transition of the fixed-connectivity surface model. Therefore, we firstly in the next section compare the results of the PTMC technique with those of the Flat histogram MC technique (FHMC) and MMC, both of which are successfully used to simulate the first order transition of the fixed-connectivity surface model \cite{KOIB-PRE-2005,KOIB-JSTP-2010}.

\section{Results}
\subsection{Accuracy and speed of PTMC}
\begin{figure}[tb]
\centering
\includegraphics[width=10.5cm]{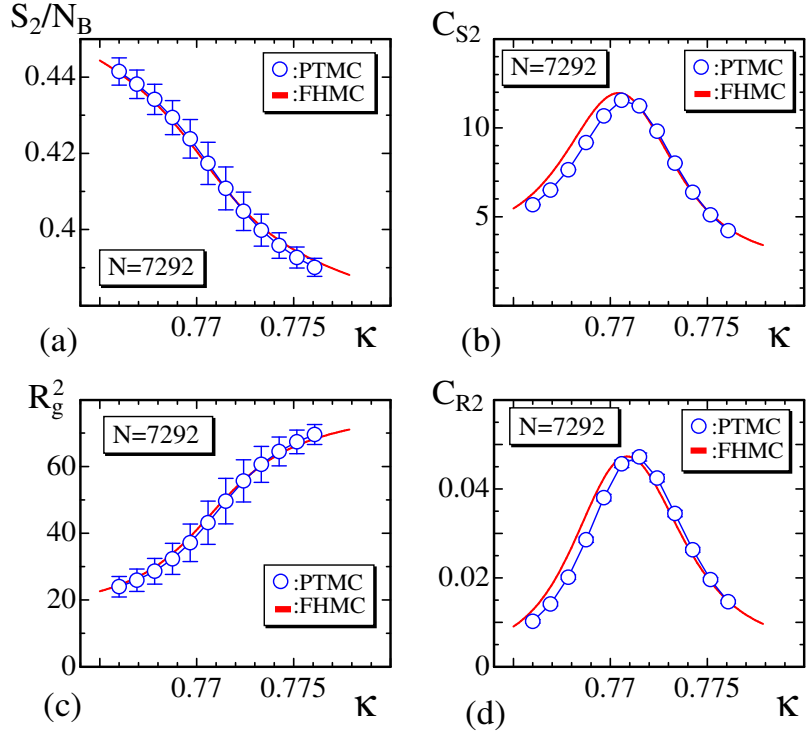}
\caption{(a) The bending energy $S_2/N_B$ vs. $\kappa$, (b) the specific heat $C_{S_2}$ vs. $\kappa$, (c) the mean square radius of gyration $R_g^2$ vs. $\kappa$, and (d) the variance $C_{R^2}$ vs. $\kappa$. The solid curves are the results of FHMC \cite{KOIB-JSTP-2010}.}
\label{fig-1}
\end{figure}
We firstly study the accuracy, speed and effectiveness of the PTMC simulation. 
The mean square radius of gyration $R^2$ is defined by
\begin{eqnarray}
\label{R2}
R_g^2=\frac{1}{N}\sum_i\left({\bf r}_i-\bar{\bf r}\right)^2, \quad
\bar{\bf r}=\frac{1}{N}\sum_i {\bf r}_i,
\end{eqnarray}
where $\bar{\bf r}$ is the center of the mass of surface. 
The $R_g^2$  can reflect the surface size because of its definition. Indeed,  $R_g^2$  becomes large (small) when the surface is swollen (collapsed). The specific heat $C_{S_2}$, which is the variance of the bending energy $S_2$,
and the variance $C_{R^2}$ of $R_g^2$ are given by 
\begin{eqnarray}
\label{CR2}
C_{S_2}=\frac{\kappa^2}{N}\left(\langle S_2^2\rangle-\langle S_2\rangle^2\right), \quad  C_{R^2}=\frac{1}{N}\left(\langle(R_g^2)^2\rangle-\langle R_g^2\rangle^2\right).
\end{eqnarray}
The $C_{R^2}$ has a peak at the phase transition point $\kappa_c$ because $R_g^2$ is expected to fluctuate at the transition point. The $C_{S_2}$ also has a peak at $\kappa_c$, and therefore  the phase transition is reflected in both $C_{R^2}$ and $C_{S_2}$ as an anomalous peak. 

In Figs. \ref{fig-1}(a),(b),  $S_2/N_B$ and $C_{S_2}$ are plotted  and compared with those of FHMC. The total number of MCS for MMC in the process (i) is fixed to $N_s\!=\!5$, and the total number of MCS for PTMC is $1\times 10^8$. 
The results of PTMC ({\color{blue}$\bigcirc$}) are almost identical with those of FHMC (solid line). The peak values of $C_{S_2}$ and $C_{R^2}$ of PTMC are also identical with those obtained by the FHMC technique. This implies that the PTMC can be used as a technique for the membrane simulations.

We comment on the speed of PTMC using Intel Fortran with OpenMP interface \cite{Intel}. The speed of PTMC with OpenMP is at least three times faster than PTMC without OpenMP on 6 cores CPU of $N_{\rm th}\!=\!12$, where $N_{\rm th}$ is the total number of threads. The simulations are performed on $N\!=\!4842$ lattice, and $N_s\!=\!20$ for the update process (i). The total number of replicas $N_R$ is fixed to $N_R\!=\!12$, which is identical with $N_{\rm th}$. In general, $N_R$ should be one of multiples of $N_{\rm th}$. 

\subsection{First order transition of the fixed-connectivity model}
\begin{figure}[hbt]
\centering
\includegraphics[width=10.5cm]{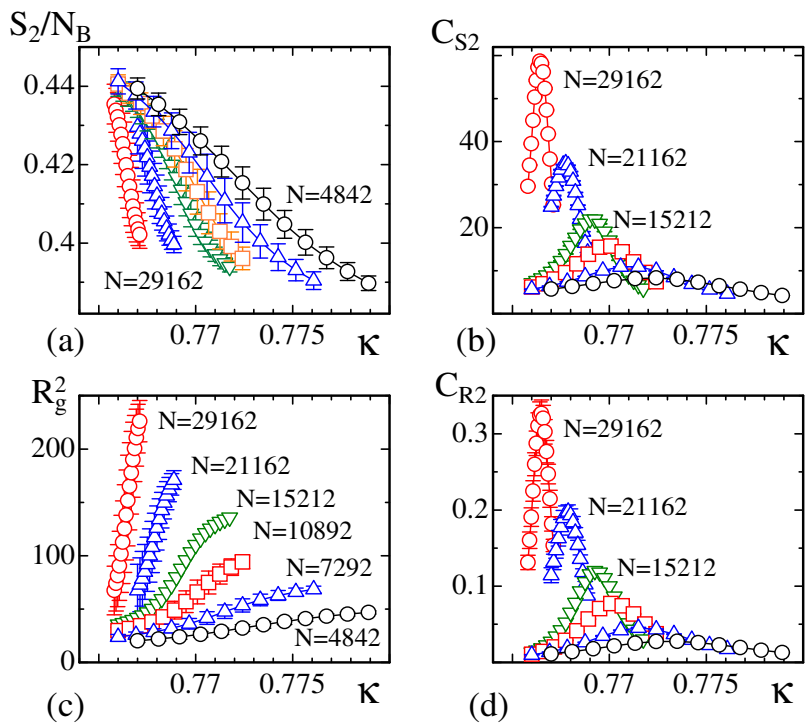}
  \caption{(a) The bending energy $S_2/N_B$ vs. $\kappa$, (b) the specific heat $C_{S_2}$ vs. $\kappa$, (c) the mean square gyration $R_g^2$  vs. $\kappa$, and (d) the variance $C_{R^2}$  vs. $\kappa$. The solid lines connecting the symbols are simply drawn to guide the eyes.}
\label{fig-2}
\end{figure}
In this subsection, we show the results obtained by relatively large scale simulations and compare them with the known results, which were reported also by our group \cite{KOIB-PRE-2005,KOIB-JSTP-2010}. The lattice size in this paper is increased up to $N\!=\!29162$, which is almost twice of that used in Refs. \cite{KOIB-PRE-2005,KOIB-JSTP-2010}. The total number of MCS in the PTMC process (i) is $N_s\!=\!30$ for the lattices $N\!=\!29162$, $N\!=\!21162$ and $N\!=\!15212$,  and $N_s\!=\!20$ for the lattices $N\!\leq\!10892$. The simulations are done on four-core and  8-thread CPUs. The total number of MCS for PTMC is $2.6\!\times\! 10^7\sim 3.7 \!\times\! 10^7$ for the $N\!=\!29162$, $N\!=\!21162$ and  $N\!=\!15212$ lattices, and  $14 \!\times\! 10^7\sim 16 \!\times\! 10^8$ for $N\!\leq\!10892$ and $N\!=\!7292$ lattices, and almost the same MCS for PTMC is assumed for the smaller lattices. The rate of acceptance for the process (i) is fixed about $50\%$,  and that of the process (ii) is $91\% \sim 97\%$ for all lattices. 

The bending energy $S_2/N_B$ and the specific heat $C_{S_2}$ are shown in Figs. \ref{fig-2}(a), (b). The mean radius of gyration $R_g^2$ and its variance $C_{R^2}$ are also plotted in Figs. \ref{fig-2}(c), (d). We find that the peaks of $C_{S_2}$ and  $C_{R^2}$ grow with increasing $N$. This indicates the existence of a phase transition \cite{Landau-PRB1976,Binder-Landau-PRB1984,Landau-PRB1986,DP-Landau-1990}.  

\begin{figure}[bt]
\centering
\includegraphics[width=10.5cm]{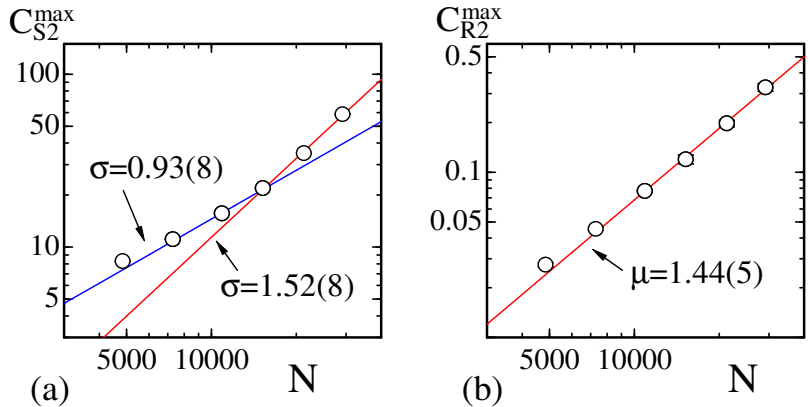}
  \caption{The log-log plots of (a)  $C_{S_2}^{\rm max}$ vs. $N$ and (b) $C_{R_g}^{\rm max}$ vs. $N$. The straight lines in  (a) are drawn by fitting the largest three data and the intermediate three data obtained on the surfaces of size $7292\!\leq\! N\!\leq\! 15212$ and $15212\!\leq\! N\!\leq\! 29162$, and the line in (b) is drawn by fitting the data for the lattices $10812\!\leq\! N\!\leq\! 29162$. }
\label{fig-3}
\end{figure}
The peak heights $C_{S_2}^{\rm max}$ and $C_{R^2}^{\rm max}$ vs. $N$ are plotted in log-log scale in Figs. \ref{fig-3}(a), (b).  The straight lines are drawn by fitting the data, and we have
\begin{eqnarray}
\label{fitting}
C_{S_2}^{\rm max} \sim N^\sigma, \quad \sigma=1.52\pm 0.08,\nonumber \\
C_{R^2}^{\rm max} \sim N^\mu,\quad \mu=1.44\pm 0.05
\end{eqnarray}
from the finite-size scaling effects \cite{Landau-PRB1976,Binder-Landau-PRB1984,Landau-PRB1986,DP-Landau-1990}. 
The exponent $\sigma\!=\!1.52(8)$ is larger than  $\sigma\!=\!0.98(12)$, which is the result of FHMC in Ref. \cite{KOIB-JSTP-2010}, and  $\sigma\!=\!0.93(13)$ of MMC in Ref. \cite{KOIB-PRE-2005}. The exponent $\mu\!=\!1.44(5)$ is also  larger than $\mu\!=\!1.24(7)$ of FHMC in \cite{KOIB-JSTP-2010}.  We should note that the exponent $\sigma$ is influenced by the size effect, because the bending energy $S_2$ reflects the surface fluctuations, which are expected to be  dominated by the long wave length modes. To compare the results of the lattices $N\!\leq\! 15212$ with that of  FHMC in \cite{KOIB-JSTP-2010}, we draw the line in  Fig. \ref{fig-2}(a) using the data of $7292\!\leq\! N\!\leq\! 15212$, and we have   $\sigma\!=\!0.93(8)$. This value $\sigma\!=\!0.93(8)$ for $C_{S_2}^{\rm max}$ is consistent with the above mentioned results of FHMC and MMC.  We also have $\mu\!=\!1.32(6)$ using $C_{R^2}^{\rm max}$ obtained on the lattices of size  $7292\!\leq\! N\!\leq\! 15212$, this value of $\mu$ is consistent with $\mu\!=\!1.24(7)$ of FHMC in \cite{KOIB-JSTP-2010} within the error. Thus, we have confirmed that the first order transition between the smooth phase and the crumpled or collapsed phase is considerably stronger than we have expected from the reported results.  It should be noted that both $\sigma$ and $\mu$ are comparable to those of the Landau-Ginzburg model, of which the first-order transition is confirmed also by the two-phase coexistence \cite{KOIB-SHOB-IJMPC-2014}.

The size effect of first order transition is characterized by the divergence of the response function according to $N\!=\!L^D$, where $L$ is the linear extension of the system and $D$ is the spatial dimension \cite{Binder-Landau-PRB1984,Landau-PRB1986,DP-Landau-1990,note-1}. The spatial dimension $D$ of the surface including the one in this paper is always considered to be $D\!=\!2$, which is the topological dimension. If this ($D\!=\!2$) is true, the results in Eq. (\ref{fitting}) are clearly larger than the commonly believed value ($\sigma\!=\!1\!=\!\mu$) for the first order transitions. The precise reason for this is unknown at present, however, the fact that $\sigma\!>\!1, \mu\!>\!1$ is a common feature of the surface models, of which the transition accompanies a change of the fractal dimension and the surfaces are not self-avoiding \cite{KOIB-SHOB-IJMPC-2014}.  

\begin{figure}[bt]
\centering
\includegraphics[width=10.5cm]{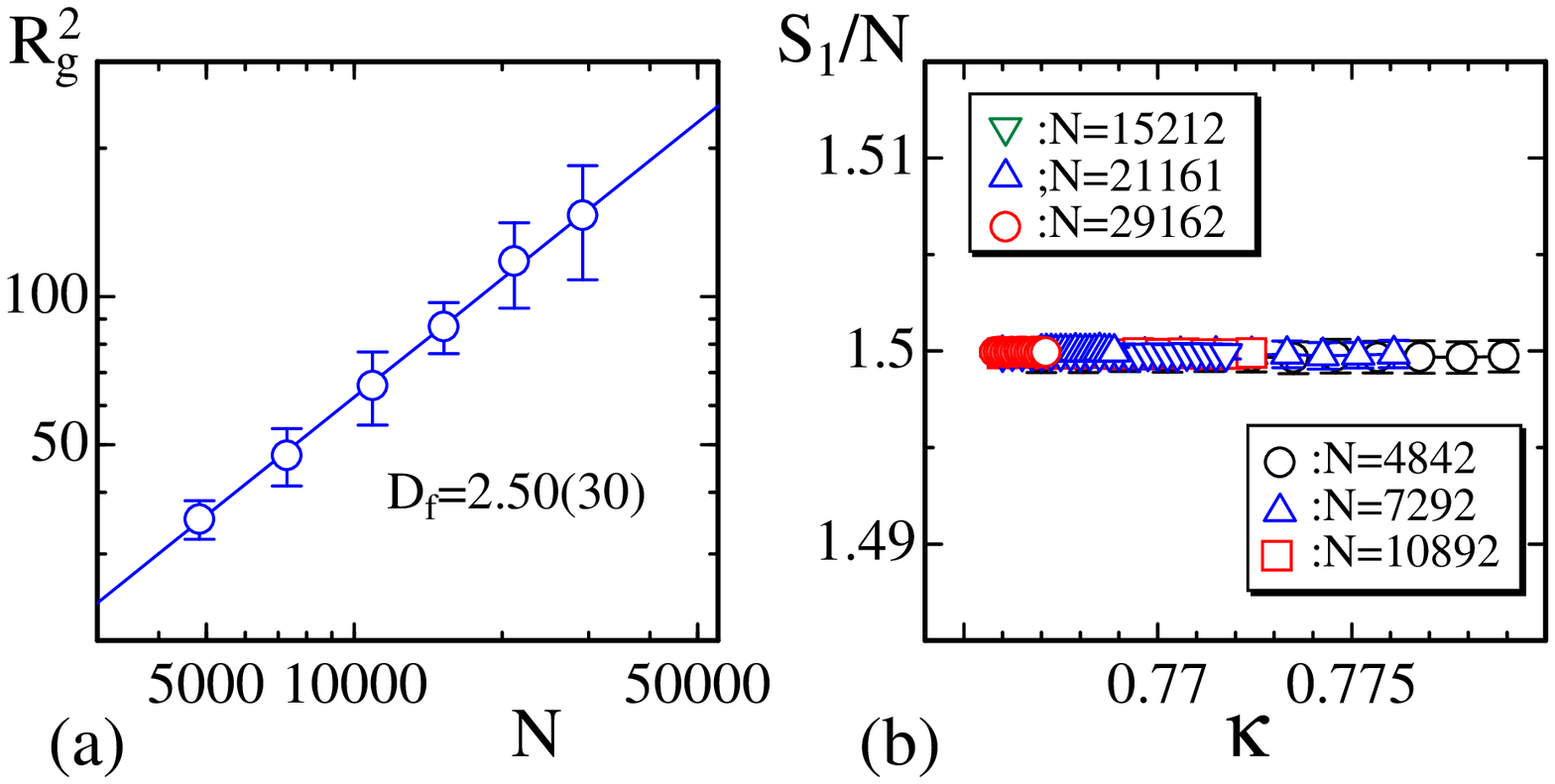}
  \caption{(a) The log-log plot of $R_g^2$ vs. $N$ at the transition points, and (b) $S_1/N$ vs. $\kappa$. The slope of the straight line in (a) gives $D_f\!=\!2.50(30)$, which is identical to the Flory estimate $D_F\!=\!2.5$. The result $S_1/N\!=\!3/2$ in (b) is consistent with the prediction from the scale invariance of $Z$.}
\label{fig-4}
\end{figure}
The mean square radius of gyration $R_g^2$ at the transition point is shown against $N$ in Fig. \ref{fig-4}(a). The straight line is drawn by fitting the data to
\begin{eqnarray}
\label{Fractal-dim}
R_g^2\sim N^{2/D_f},\quad D_f=2.50\pm 0.30,
\end{eqnarray}
where $D_f$ is the fractal dimension. This value $D_f\!=\!2.50(30)$ is considered as the fractal dimension at the transition point, although the well-defined value does not always exist for any physical quantities at the first order transition point in the limit of $N\!\to\!\infty$. We should note that the result of Eq. (\ref{Fractal-dim}) is identical to the so-called Flory estimate $D_F\!=\!2.5$ \cite{NELSON-SMMS2004-Gompper,Bowick-PREP2001}. The Gaussian bond potential (Fig. \ref{fig-4}(b)) also satisfies the expected values $S_1/N\!=\!3/2$, which comes from the scale invariance of $Z$ \cite{WHEATER-JP1994}. This result supports that the simulations are correctly performed.  

\begin{figure}[bt]
\centering
\includegraphics[width=10.5cm]{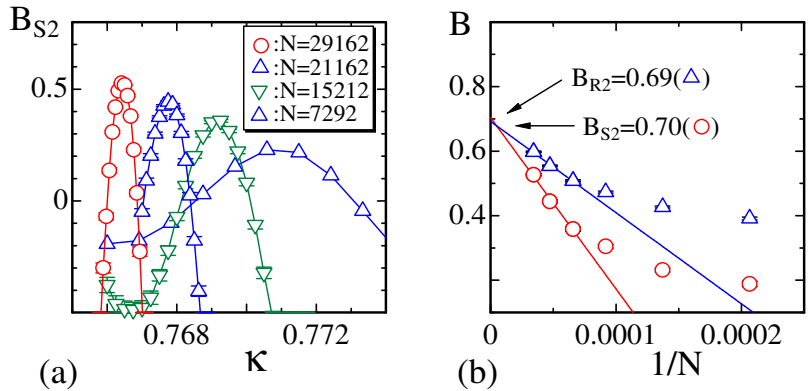}
  \caption{(a) The Binder quantity $B_{S_2}$ vs. $\kappa$,  and (b) the linear extrapolations of $B_{S_2}$ and  $B_{R^2}$ vs. $1/N$ obtained at the transition point of the surface of size $4842\!\leq\! N\!\leq\!29162$. The straight lines in (b) are drawn by using the data of the largest three  lattices.}
\label{fig-5}
\end{figure}
The order of the transitions can also be characterized by the Binder quantity $B_{S_2}$ \cite{BINDER-ZFPB-1981}, which is defined by
\begin{equation}
\label{Binder-cumulants}
B_{S_2}=1-{\langle \left(S_2-\langle S_2\rangle\right)^4 \rangle \over 3\langle \left(S_2-\langle S_2\rangle\right)^2 \rangle^2 }.
\end{equation}
If the transition is first order, it is expected that $B\!=\!2/3$. In Fig.\ref{fig-5}(a) we plot $B_{S_2}$, which has the peak at the transition point for each $N$. To get $B_{S_2}(N\!\to\!\infty)$, the peak value of $B_{S_2}$ is plotted against $1/N$ in Fig.\ref{fig-5}(b). By the linear extrapolation, we have $B_{S_2}\!=\!0.70$, and we also have  $B_{R^2}\!=\!0.69$ by using the same technique. Both results are consistent with the fact that the model undergoes the first order transition.

\begin{figure}[bt]
\centering
\includegraphics[width=10.5cm]{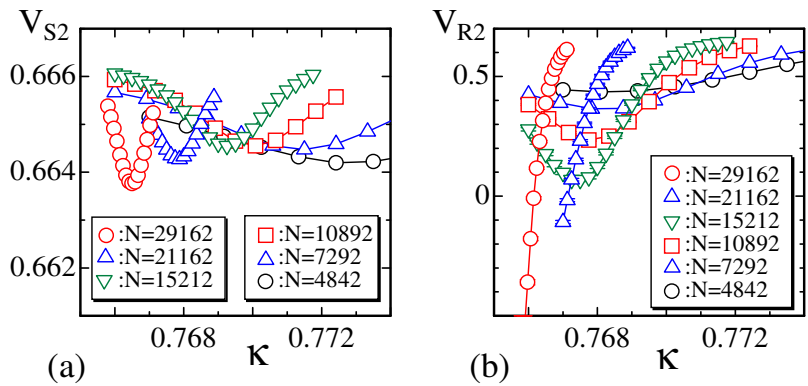}
  \caption{The Binder cumulants (a) $V_{S_2}$ vs. $\kappa$,  and (b)  $V_{R^2}$ vs. $\kappa$.  The minimum $V_{\rm min}$ becomes apparent with increasing $N$ in both quantities. The error bars for $V_{S_2}$ are not shown to clarify the data. }
\label{fig-6}
\end{figure}
The first order transition can also be reflected in the Binder cumulant
\begin{equation}
\label{Binder-cumulant-S2}
V_{S_2}=1-{\langle S_2^4 \rangle \over 3\langle S_2^2 \rangle^2 }.
\end{equation}
In Ref. \cite{Landau-PRB1986}, the quantity denoted by $V_L$ for the $q$-state Potts model is shown to behave quite differently at first and second order transitions. For 10-state Potts model, which has the first-order transition, $V_L$ has a minimum $V_{\rm min}$ and this $V_{\rm min}$ becomes apparent as $N$ increases. On the contrary, for 2-state (Ising model) and 3-state Potts models, which have continuous transitions, the minimum $V_{\rm min}$ also appears in $V_L$, however it disappears for sufficiently large $N$. 
In Figs. \ref{fig-6}(a),(b), $V_{S_2}$ and $V_{R^2}$ are plotted, where $V_{R^2}$ is defined by replacing $S_2$ with $R_g^2$ in Eq. (\ref{Binder-cumulant-S2}). Since $R_g^2$ has a double peak structure, just like $S_2$, in its probability distribution at the transition point \cite{KOIB-PRE-2005,KOIB-JSTP-2010,KOIB-SHOB-IJMPC-2014},  $V_{R^2}$ is also expected to be sensitive to the order of the transition. Indeed, we see the expected behavior in  both $V_{S_2}$ and $V_{R^2}$ in  Figs. \ref{fig-6}(a),(b). We should note that the position $\kappa$ of $V_{R^2}^{\rm min}$ is slightly different from (or moves to the left of) that of $V_{S_2}^{\rm min}$. On the largest lattices of $N\!=\!21162$ and $N\!=\!29162$,  the position of $V_{R^2}^{\rm min}$ is expected to be less than the smallest $\kappa$ of the data range, and therefore $V_{R^2}^{\rm min}$ does not appear in the plots.  It should also be remarked that $V_{S_2}^{\rm min}$ increases at least for $N\!\leq\!10892$ with increasing $N$ and starts to decrease on the lattices of $N\!=\!21162$. This implies that large lattices are necessary to get the proper finite-size effect of $S_2$.

\section{Summary and conclusions}
In this study, the canonical surface model for membranes is simulated by means of the PTMC technique. The PTMC simulations, which are parallelized with OpenMP, are performed on the multi-core CPUs. The results are compared with those previously obtained by the Metropolis MC and the flat histogram MC simulations. The Binder quantities are also calculated to confirm that the model undergoes a first order phase transition. To summarize the results:
\begin{enumerate}
\item[(1)] The PTMC technique is successfully used to simulate the canonical triangulated surface model, which is known to undergo a first order transition.
\item[(2)] The first order transition is found to be considerably stronger than we have expected from the reported results. 
\end{enumerate}
We should note that the speed of the PTMC simulation with the openMP parallelization on multi-cores CPUs is sufficiently faster than that of the single-core CPUs.

The simulations of the surface model in higher dimensions is time consuming because of the low convergence speed for the update of ${\bf r}$. Therefore, we expect that the surface model, such as the Landau-Ginzburg model, in higher dimensions is efficiently simulated by the PTMC technique. 

\vspace{0.5cm}
\noindent
{\bf Acknowledgment}

\noindent
We are grateful to prof. Hideo Sekino for the support of the Promotion of Joint Research 2014, Toyohashi University of Technology. This work is supported in part by the Grant-in-Aid for Scientific Research (C) Number 26390138.





\begin{thebibliography}{00}





\bibitem{NELSON-SMMS2004-Leibler}
S. Leibler, \textit{Equilibrium statistical mechanics of fluctuating films and membranes}, in \textit{Statistical Mechanics of Membranes and Surfaces, Second Edition}, edited by  D. Nelson, T.Piran, and S.Weinberg, (World Scientific, 2004), p.49.

\bibitem{NELSON-SMMS2004-Gompper}
G. Gompper and D.M. Kroll, \textit{Triangulated-surface models of fluctuating membranes}, in \textit{Statistical Mechanics of Membranes and Surfaces, Second Edition}, edited by  D. Nelson, T.Piran, and  S.Weinberg, (World Scientific, 2004), p.359.

\bibitem{Bowick-PREP2001}
 M. Bowick and A. Travesset, Phys. Rep. {\bf 344}, 255 (2001).

\bibitem{Wiese-PTCP2000}
K. Wiese,  \textit{Polymerized Membranes, a Review}, in \textit{Phase Transitions and Critical Phenomena}, Vol. 19, C.Domb, J.Lebowitz  (Eds.),  (Academic Press, London, 2000), p.253.

\bibitem{KANTOR-NELSON-PRA1987}
 Y. Kantor and  D.R. Nelson, Phys. Rev. A {\bf 36}, 4020 (1987).

\bibitem{Peliti-Leibler-PRL1985}
 L. Peliti and S. Leibler, Phys. Rev. Lett. {\bf 54}, (15) 1690  (1985).

\bibitem{DavidGuitter-EPL1988}
 F. David and E. Guitter, Europhys. Lett,  {\bf 5}, (8) 709  (1988).

\bibitem{PKN-PRL1988}
M. Paczuski, M. Kardar, and D. R. Nelson, Phys. Rev. Lett. {\bf 60}, 2638 (1988).

\bibitem{BKS-PLA2000}
 M.E.S. Borelli, H. Kleinert, and Adriaan M.J. Schakel, Phys. Lett. A {\bf 267}, 201 (2000).

\bibitem{Kownacki-Mouhanna-2009PRE}
J.-P. Kownacki and D. Mouhanna, Phys. Rev. E. {\bf 79}, 040101 (2009).

\bibitem{Essa-Kow-Mouh-PRE2014}
K. Essafi, J.-P. Kownacki, and D. Mouhanna, Phys. Rev. E {\bf 89}, 042101 (2014).

\bibitem{KD-PRE2002}
J-P. Kownacki and H. T. Diep, Phys. Rev. E {\bf 66}, 066105 (2002).

\bibitem{KOIB-PRE-2005}
H. Koibuchi and T. Kuwahata, Phys. Rev. E {\bf 72}, 026124 (2005).

\bibitem{KOIB-JSTP-2010}
H. Koibuchi, J. Stat. Phys. {\bf 140}, 676 (2010).

\bibitem{HELFRICH-1973}
 W. Helfrich, Z. Naturforsch, {\bf 28}c, 693 (1973).

\bibitem{POLYAKOV-NPB1986}
 A.M. Polyakov, Nucl. Phys. B {\bf 268}, 406 (1986);\\
H. Kleinert, Phys. Lett. B {\bf 174}, 335 (1986).

\bibitem{KOIB-SHOB-IJMPC-2014}
H. Koibuchi, Int. J. Mod. Phys. C. {\bf 25}, 145033 (2014).

\bibitem{Mepropolis-JCP-1953}
N. Metropolis, A. W. Rosenbluth, M. N. Rosenbluth and
A. H. Teller, J. Chem. Phys. {\bf 21}, 1087 (1953).

\bibitem{Landau-PRB1976}
D.P. Landau,  Phys. Rev. B {\bf 13}, 2997 (1976).

\bibitem{Karin-Taylor-1975}
S. Karlin and H. Taylor,  \textit{A First Course in Stochastic Processes, Second Edition}, (Academic Press, 1975). 

\bibitem{A-Ueda-1990}
A. Ueda, \textit{Computer Simulations: Atomistic Motion in Macroscopic Systems},  (in Japanese), (Asakura Shoten, 1990).

\bibitem{Hukushima-Nemoto-JPSOC1996}
K. Hukushima and K. Nemoto, J. Phys. Soc. Jpn. {\bf 65}, 1604 (1996). 

\bibitem{Takayama-Hukushima-JPSOC2007}
H. Takayama and K. Hukushima, J. Phys. Soc. Jpn. {\bf 76}, 013702 (2007).

\bibitem{New-Mag-Hans-PRE2007}
T. Neuhaus, M. P. Magiera, and U. H. E. Hansmann,
Phys. Rev. E {\bf 76}, 045701(R) (2007).

\bibitem{C-E-Fios-PRE2008}
Carlos E. Fiore, Phys. Rev. E {\bf 78}, 041109 (2008).

\bibitem{Intel}
See documents for Intel Parallel Studio$^{\textregistered}$, for example.

\bibitem{AMBJORN-NPB1993}
J. Ambjorn, A. Irback, J. Jurkiewicz, and B. Petersson, Nucl. Phys. B {\bf 393}, 571 (1993).

\bibitem{WHEATER-JP1994}
J.F. Wheater, J. Phys. A Math. Gen. {\bf 27}, 3323 (1994).

\bibitem{Binder-Landau-PRB1984}
K. Binder and D.P. Landau,  Phys. Rev. B {\bf 30}, 1477 (1984).

\bibitem{Landau-PRB1986}
Murty S. S. Challa, D.P. Landau and K. Binder,  Phys. Rev. B {\bf 34}, 1841 (1986).

\bibitem{DP-Landau-1990}
D.P. Landau,  \textit{Monte Carlo studies of finite size effects at first and second order phase transitions}, in \textit{Finite size scaling and numerical simulation of statistical systems}, V. Privman (Ed.), pp.225-260, (World Scientific, 1990). 

\bibitem{note-1}
The symbol $D$ is used here for the spatial dimension instead of $d$  to distinguish it from the dimension $d(=\!3)$ of the external space ${\bf R}^d$.

\bibitem{BINDER-ZFPB-1981}
K. Binder, Z. Phys. B \textbf{43}, 119 - 140 (1981).

\end{thebibliography}
\end{document}